\documentclass[useAMS,usenatbib]{mn2e}

\usepackage{epsfig}
\usepackage{amssymb}
\usepackage{amsmath}
\newcommand{\mdot}{M$_{\odot}$}

\usepackage{color}

\defcitealias{Emonts2005}{EM05}

\usepackage{mathptmx}

\title[Jet-driven outflows of ionised gas in 3C293]{Jet-driven outflows of ionised gas in the nearby radio galaxy 3C293}
\author[E. K. Mahony et al.]{E. K. Mahony$^{1}$\thanks{E-mail:mahony@astron.nl},  J. B. R. Oonk$^{1,2}$,  R. Morganti$^{1,3}$, C. Tadhunter$^{4}$, P. Bessiere$^{5}$,\and P. Short$^{4}$, B. H. C. Emonts$^{6}$ and T. A. Oosterloo$^{1,3}$ \\
$^{1}$ASTRON, the Netherlands Institute for Radio Astronomy, Postbus 2, 7990 AA, Dwingeloo, The Netherlands.\\
$^{2}$Leiden Observatory, Leiden University, P.O. Box 9513, 2300 RA Leiden, The Netherlands.\\
$^{3}$Kapteyn Astronomical Institute, University of Groningen, Postbus 800, 9700 AV Groningen, The Netherlands.\\
$^{4}$Department of Physics \& Astronomy, University of Sheffield, Sheffield S3 7RH. \\
$^{5}$Universidad de Concepci{\'o}n, Departamento de Astronomi{\'a}, Concepci{\'o}n, Chile. \\
$^{6}$Centro de Astrobiolog\'ia (INTA-CSIC), Ctra de Torrej\'on a Ajalvir, km 4, 28850 Torrej\'on de Ardoz, Madrid, Spain 
}

\begin{document}

\date{Accepted 2015 .... Received 2015 ...; in original form 2015 ...}

\pagerange{\pageref{firstpage}--\pageref{lastpage}} \pubyear{2015}

\maketitle

\label{firstpage}

\begin{abstract}

Fast outflows of gas, driven by the interaction between the radio-jets and ISM of the host galaxy, are being observed in an increasing number of galaxies. One such example is the nearby radio galaxy 3C293. In this paper we present Integral Field Unit (IFU) observations taken with OASIS on the William Herschel Telescope (WHT), enabling us to map the spatial extent of the ionised gas outflows across the central regions of the galaxy. The jet-driven outflow in 3C293 is detected along the inner radio lobes with a mass outflow rate ranging from $\sim 0.05-0.17$ \mdot\,yr$^{-1}$ (in ionised gas) and corresponding kinetic power of $\sim 0.5-3.5\times 10^{40}$\,erg\,s$^{-1}$. Investigating the kinematics of the gas surrounding the radio jets (i.e. not directly associated with the outflow), we find line-widths broader than $300$ km\,s$^{-1}$ up to 5\,kpc in the radial direction from the nucleus (corresponding to 3.5\,kpc in the direction perpendicular to the radio axis at maximum extent). Along the axis of the radio jet line-widths $>400$ km\,s$^{-1}$ are detected out to 7\,kpc from the nucleus and line-widths of $>500$ km\,s$^{-1}$ at a distance of 12 kpc from the nucleus, indicating that the disturbed kinematics clearly extend well beyond the high surface brightness radio structures of the jets. This is suggestive of the cocoon structure seen in simulations of jet-ISM interaction and implies that the radio jets are capable of disturbing the gas throughout the central regions of the host galaxy in all directions.

\end{abstract}


\section{Introduction}

In order to reproduce the tight correlations observed between the supermassive black hole (SMBH) and the host galaxy, AGN driven outflows have become a key ingredient in galaxy formation and evolution models (see e.g. \citealt{Silk1998, DiMatteo2005, croton06, bower06, Ciotti2010, Fabian2012}). Fast outflows are generally invoked to regulate the growth of the galaxy by halting the accretion of gas onto the central AGN and prohibiting star formation. There are a number of physical mechanisms thought to be resposible for driving these outflows such as starburst driven winds (see review by \citealt{Veilleux2005}) or radiation pressure from the accretion disk (e.g. \citealt{King2005,Zubovas2014a}). However, there is increasing evidence that in radio loud AGN the interaction between the radio jets and ISM of the host galaxy can also be responsible for large outflows. Massive and fast outflows of neutral or molecular gas driven by the radio jets have been observed in a growing number of objects \citep{Morganti2005, Lehnert2011, Dasyra2012, Morganti2013a, McNamara2014, Morganti2015}.
 
In the optical, jet-induced outflows of ionised gas have also been observed \citep{Emonts2005, Morganti2005, Morganti2007, Holt2010}. While many of these studies used long-slit observations, the increasing availability of Integral Field Units (IFUs) means we can now trace the impact of these outflows in 2D \citep{Davis2012,Harrison2014}. Pinpointing the location and spatial extent of these outflows enables us to derive crucial parameters, such as the mass outflow rates and kinetic energy involved, which we can compare to predictions from galaxy evolution simulations. Current theoretical models of jet-ISM interaction \citep{Wagner2011, Wagner2012} predict that these fast outflows are produced by shocks, which ionise the surrounding gas. As the gas then cools in the post-shock region, outflows in HI and CO are observed. Determining whether ionised and neutral gas outflows originate from the same location in the galaxy will therefore provide important constraints for these models.

One such example of jet-induced outflows is observed in 3C293, a nearby radio galaxy at a redshift of $z=0.045$ \citep{Sandage}. 3C293 is a `double-double' radio galaxy, where the older, outer lobes have a projected linear size of $\sim$190\,kpc and the younger, inner radio lobes are about 1.7 kpc across \citep{Akujor1996, Joshi2011, Beswick2004}. Fast outflows in 3C293, up to 1400\,km\,s$^{-1}$, were first detected in neutral hydrogen using the Westerbork Synthesis Radio Telescope \citep{Morganti2003}. However, the spatial resolution was not high enough to determine where along the inner $\sim$1\,kpc jets the outflows originated from. To investigate this further, long-slit spectroscopy was taken along the radio jet axis where outflows of ionised gas were detected associated with the eastern radio lobe. The profile of the HI outflows and the ionised gas outflows were remarkably similar and it was therefore concluded that the HI outflows were associated with the same jet-ISM interaction \citep[hereafter EM05]{Emonts2005}.

Recent observations using the Karl G. Jansky Very Large Array (VLA) provided higher spatial resolution data needed to localise the HI outflows on arcsecond scales. Unexpectedly, these data show that the HI outflows are coming from the western radio lobe, opposite to where the most extreme ionised outflows are detected \citep{Mahony2013}. 

In addition to the detection of outflows of cool gas, 3C293 also contains a large amount of warm molecular hydrogen, possibly heated by shocks driven by the interaction between the radio jets and the ISM \citep{Ogle2010}. To confirm this, follow-up X-ray observations revealed 10$^7$\,K shock-heated gas in both the nucleus and the inner radio jets, again driven by jet-induced shocks \citep{Lanz2015}. Studies of the cold molecular gas in this galaxy indicate a cold H$_2$ mass of 2.2$\times 10^{10}$\,\mdot and a starformation rate of 4\,\mdot\,yr$^{-1}$ \citep{Evans1999, Labiano2014}. The outflows detected in HI and ionised gas have not yet been detected in CO, however, \citet{Labiano2014} derive a 3$\sigma$ upper limit of the molecular gas outflow of 3.2\% of the total H2 mass. This limit can rule out an extreme molecular outflow like that observed in Mrk 231 \citep{Feruglio2010}, but it is within the range observed in other objects with molecular outflows (e.g. \citealt{Alatalo2011, Morganti2015}).

In this paper, we present IFU observations of 3C293 in order to confirm the location of the fast outflow of ionised gas and unambiguously determine its location with respect to the fast outflow seen in HI. IFU observations allow us, for the first time, to map the properties of the ionised gas outflow across the inner regions of 3C293. Section \ref{obs} describes the observations and data reduction and the results are presented in Section \ref{results}. In Section \ref{outflow} we discuss the properties of the outflow and how these new observations compare with previous results before concluding in Section \ref{concl}.

\section{Observations and data reduction} \label{obs}

\begin{figure}
\centering{\epsfig{file=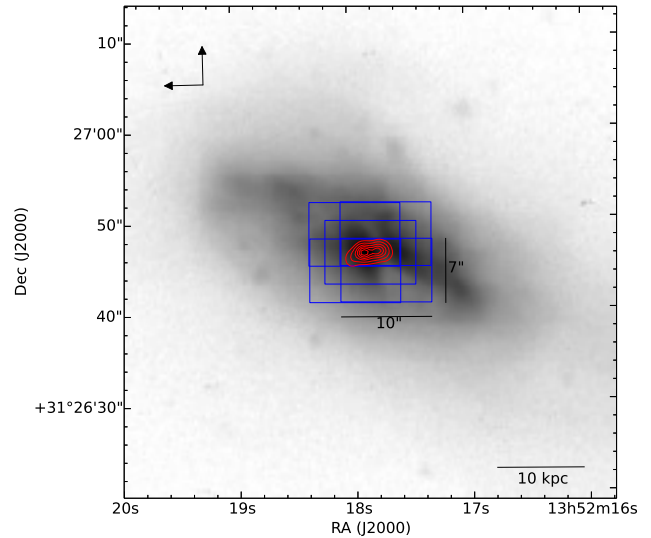, width=\linewidth}}
\caption{Optical V-band image of 3C293 observed with the Hiltner 2.4m telescope of the Michigan-Dartmouth-MIT (MDM) observatory (Emonts et al., in prep). The blue boxes show the position of each exposure; two were centred on the nucleus of 3C293 and 4 exposures were offset by 2 arcsec in a rectangular pattern. The red contours show the inner radio lobes observed at 1.4\,GHz with the Karl G. Jansky Very Large Array (VLA). The contour levels range from 0.1--1.25\,Jy in steps of 0.25\,Jy. North is up and East to the left as indicated by the arrows in the top left corner. \label{overlay}}
\end{figure}

Observations were carried out using the OASIS Integral Field Unit (IFU) on the Willam Herschel Telescope (WHT) on 09 April, 2013. Two different spectral configurations were used to target the H$\alpha$ and [NII] lines (MR661) and [SII] lines (MR735) separately. For each configuration, six exposures of 900s were taken with four of the exposures offset by 2 arcsec in a rectangular pattern. Figure \ref{overlay} shows the region covered by the OASIS observations (the blue boxes show the OASIS 10x7 arcsec field of view) overlaid on an optical V-band image (Emonts et al., in prep) with the 1.4 GHz radio contours of the inner radio jet shown in red. This exposure pattern was chosen to slightly expand the area observed whilst still maintaining the highest S/N for the central 8x3 arcsec region, coincident with the radio jets. Since the angular size of 3C293 is much larger than the field of view (f.o.v), each exposure was followed by a separate sky exposure of the same duration (900s) in order to perform adequate sky subtraction. The average seeing throughout the observations was 0.8 arcsec. 

The majority of the data reduction was carried out using the XOASIS data reduction pipeline with the exception of the sky subtraction, flux calibration and mosaicing which were done using standard procedures in IDL. The data were then resampled to 0.3 arcsec pixels and continuum subtracted. In order to detect the faint, broad signatures indicative of an outflow, the data was binned using a Voronoi binning algorithm \citep{voronoi} to ensure there was a minimum S/N ratio of 10 for every bin. This results in bin sizes which vary from $0.3\times0.3$ arcsec in the central regions to $3.6\times2.7$ arcsec in the outer regions. 

In order to remove the effects of instrumental broadening, the widths of the sky lines were measured prior to the sky subtraction and this value (6.8\AA  for MR661 and 7.1\AA  for MR735) used to deconvolve the measured line-widths. The astrometry was checked by crossmatching with 2MASS images \citep{2mass} and we estimate that the positional error between the radio and optical data is less than 0.5 arcsec. 

\subsection{Gaussian fitting}

The binned, continuum-subtracted cubes of each grism were stitched together and run through a gaussian line-fitting routine\footnote{This used the python mpfit package written by Mark Rivers (http://cars.uchicago.edu/software/python/mpfit.html) which was based on the IDL mpfit package \citep{mpfit}.}. Combining the data allowed us to fit the H$\alpha$, [NII] and [SII] lines simultaneously providing better constraints on the fitting. The line-widths of all the lines were required to be the same and the ratios of the two [SII] lines were required to fall between the high and low density limits (i.e. 0.44--1.44).

\begin{figure}
\includegraphics[width=\linewidth]{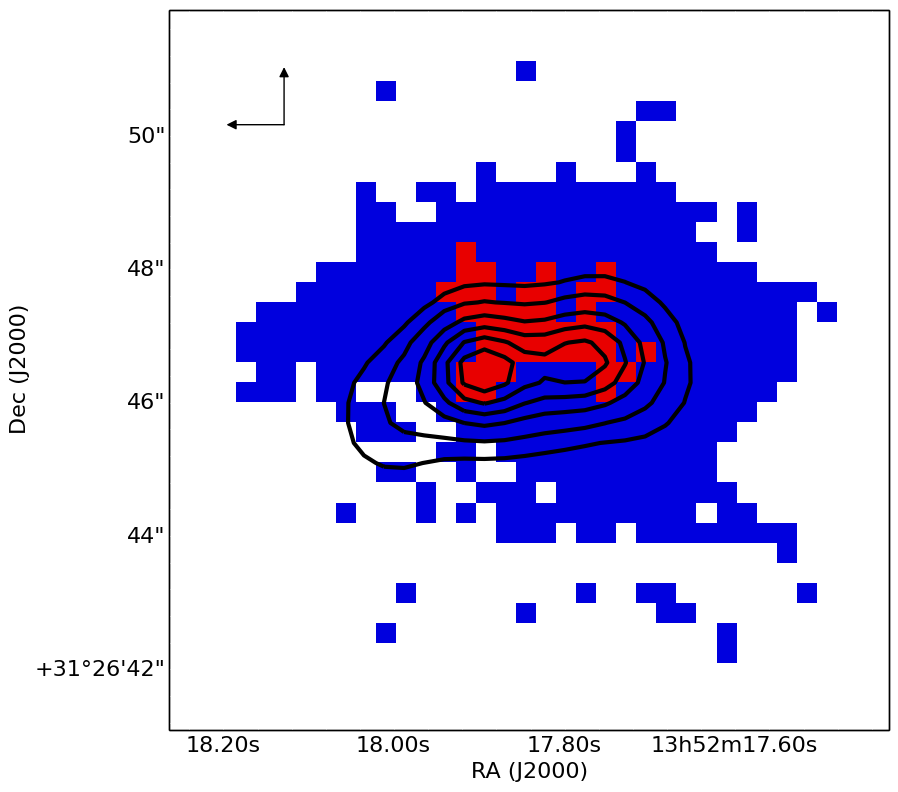}
\caption{The number of gaussian components needed to fit the data; the blue bins represent pixels where a single gaussian component was sufficient to fit the data and the red denotes regions where 2 gaussian components were required. The black contours show the inner radio lobes observed at 1.4\,GHz with the VLA \citep{Mahony2013}. The contour levels range from 0.1--1.25\,Jy in steps of 0.25\,Jy and are the same contours shown in all the following figures. North is up and East to the left as indicated by the arrows in the top left corner. \label{comps}}
\end{figure}

\begin{figure*}
\centering{\epsfig{file=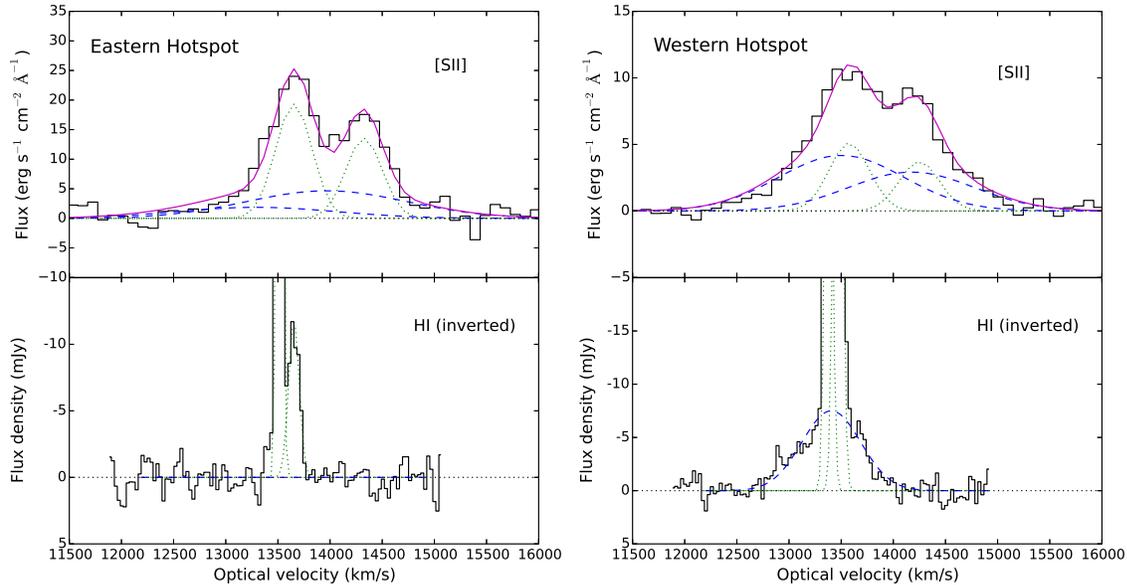, width=\linewidth}}
\caption{Example spectra extracted from regions associated with the two hotspots in 3C293. The top row shows the [SII]$\lambda$6716 and [SII]$\lambda$6737 lines observed with OASIS (the [SII]$\lambda$6716 line was used as the rest wavelength when converting to velocity). The observed data is denoted by the black histogram while the solid magenta line shows the total of the gaussian components fit to the data. In both instances a second, broader gaussian component (shown by the dashed blue line) was required to fit the broad wings observed. The bottom row shows the corresponding HI absorption spectrum (inverted for comparison) from previous VLA observations \citep{Mahony2013}. \label{examplespectra}} 
\end{figure*}

Many of the spectra exhibited broad wings that were not well fit by a single gaussian component. In these cases, an additional gaussian component was added to the model. Figure \ref{comps} shows the regions in which a 2-gaussian fit was required and some example spectra are shown in Figure \ref{examplespectra}.

\section{Mapping the ionised gas outflow in 3C293} \label{results}

In regions where 2 gaussian components were needed to fit the data, the second gaussian component was much broader than the first component fitted. As such, the two components were separated to distinguish between the gas associated with the galaxy disk, what we term the `narrow' component, and the `broad' component indicative of more extreme kinematics. We treat these two components separately in the following analysis. 

\subsection{Ionised gas kinematics} 
 
\begin{figure*}
\hspace{-1cm}
\begin{minipage}{0.45\linewidth}
\includegraphics[width=\linewidth]{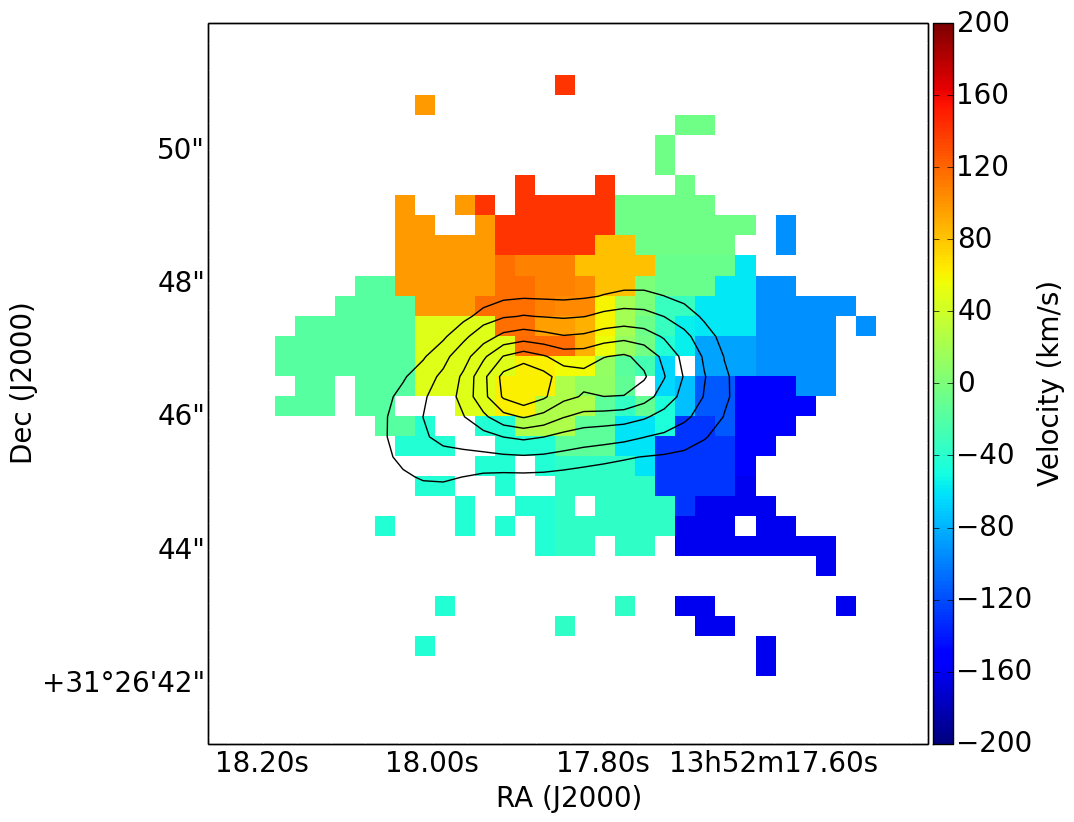}
\end{minipage}
\hspace{0.5cm}
\begin{minipage}{0.45\linewidth}
\includegraphics[width=\linewidth]{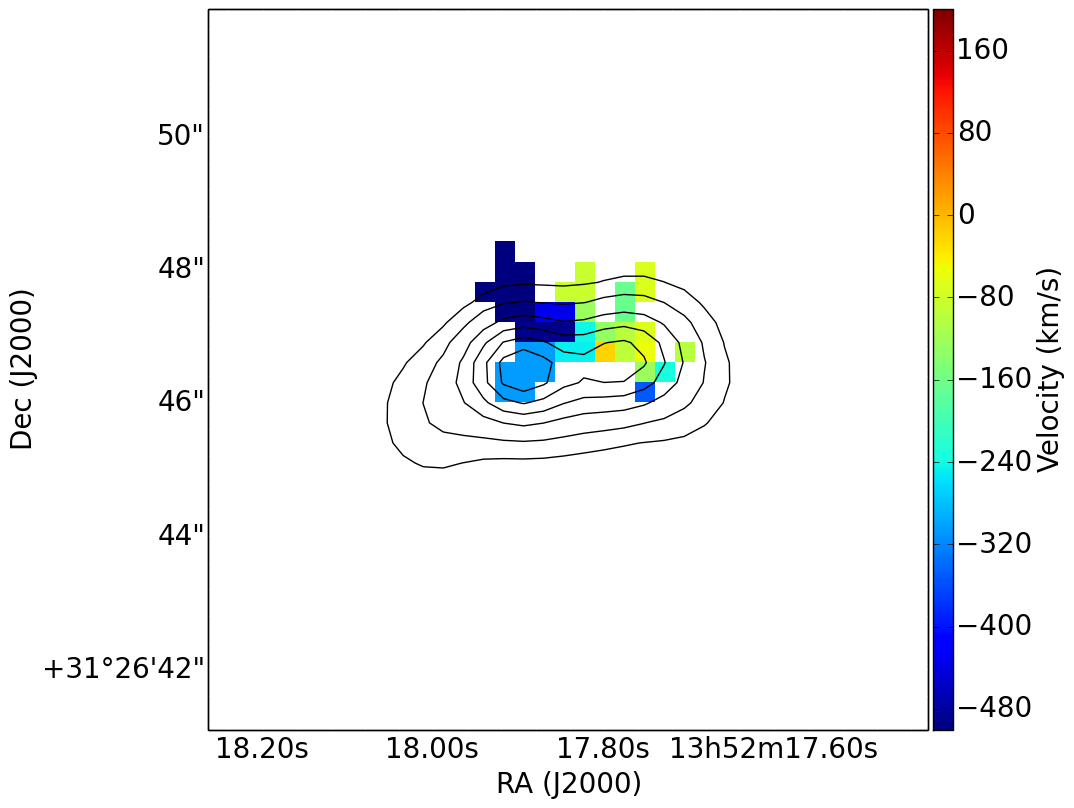}
\end{minipage}
\caption{Velocity maps of the narrow gaussian components, representing the gas associated with the disk of the galaxy (left) and broad components, representative of the outflow (right). The broad component associated with the western jet is close to the systemic velocity in this region, similar to the central velocity of the HI outflow which is also detected here \citep{Mahony2013}. At the eastern jet the broad component is significantly blueshifted, in agreement with previous long-slit observations \citepalias{Emonts2005}. For the colour plots we refer the reader to the online version of this journal.\label{velmaps}} 
\end{figure*}

Figure \ref{velmaps} shows the velocity maps for the narrow components in the left-hand panel, and the broad components on the right. The velocity map of the narrow components shows a regularly rotating disk in agreement with previous studies (\citealt{Haschick1985, Evans1999, Beswick2004}, \citetalias{Emonts2005}, \citealt{Labiano2014}), providing confidence that the gaussian line-fitting routine produces sensible fits to the data. 

Focusing now on the broad components (right panel of Figure \ref{velmaps}) we see that the gas is significantly blueshifted (with outflow velocities ranging from -300 to -500\,km\,s$^{-1}$) in regions associated with the eastern radio jet, indicative of an outflow. This is in agreement with the results obtained from the previous long-slit observations of 3C293 where a clear blueshifted outflow of ionised gas was detected at the eastern jet \citepalias{Emonts2005}. Regions coincident with the western radio jet also show evidence for blueshifted velocities, but these are less extreme than those associated with the eastern jet, with outflow velocities ranging from -50 to -150 \,km\,s$^{-1}$.

The line-widths measured for the broad components indicate very high velocity dispersions of the outflowing material all along the inner radio jet. The broadest line is detected at the eastern hotspot which has a $FWHM$ of $1755\pm141$\,km\,s$^{-1}$ while for the western jet the $FWHM$ of the outflow ranges from $1080\pm70 - 1354\pm79$\,km\,s$^{-1}$. Figure \ref{examplespectra} shows the comparison between the broad components detected in the ionised gas compared to the corresponding HI absorption spectrum associated with each hotspot. While the broad components associated with the eastern jet have larger velocity shifts and line-widths, the broad components fit to the spectrum of the western jet have a much higher peak flux. 

Given the blueshifted velocities and the broad line-widths observed, we interpret the broad component as representing a fast outflow, in line with previous studies of 3C293 (\citetalias{Emonts2005}, \citealt{Mahony2013}). These earlier investigations have led to seemingly conflicting results on where the outflow is located. The outflow detected in the neutral gas has only been detected in front of the western inner radio lobe, but previous long-slit observations have detected the most extreme ionised gas outflow associated with the eastern radio jet. As suggested in \citetalias{Emonts2005}, figures \ref{comps} and \ref{velmaps} shows that the ionised gas outflow is detected all along the radio jet axis, providing further evidence that this outflow is being driven by the radio jet. It appears that the outflow is detected preferentially in the northern regions of the radio lobes, however, this is most likely due to the higher S/N observed here. The southern regions are more obscured by the dust lane of 3C293 making it more difficult to detect the faint, broad wings. 

The line-widths of the narrow components are roughly uniform across the f.o.v, ranging from $\sim$300\,km\,s$^{-1}$ up to 450\,km\,s$^{-1}$. These values are higher than usually seen in quiescent gas, which suggests that we are tracing unsettled gas out to a few kpc from the nucleus. However, to further investigate this claim higher spectral resolution data are needed. To this effect we have compared the IFU data with higher spectral resolution long-slit data. 

\subsection{Comparison with long-slit data}  \label{longslitsec}

\begin{figure*}
\begin{minipage}{0.45\linewidth}
\centering{\epsfig{file=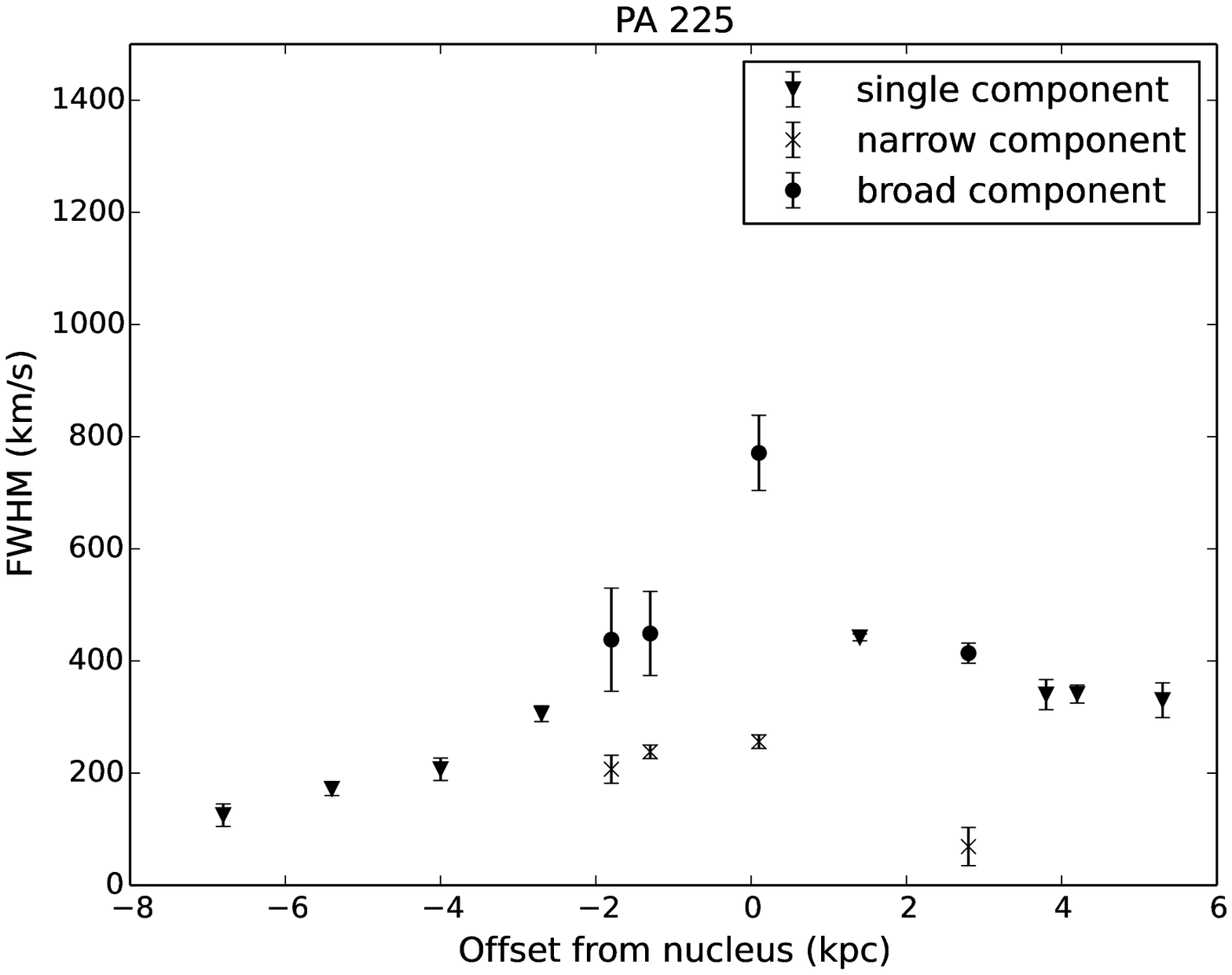, width=\linewidth}}
\end{minipage}
\hspace{0.5cm}
\begin{minipage}{0.45\linewidth}
\centering{\epsfig{file=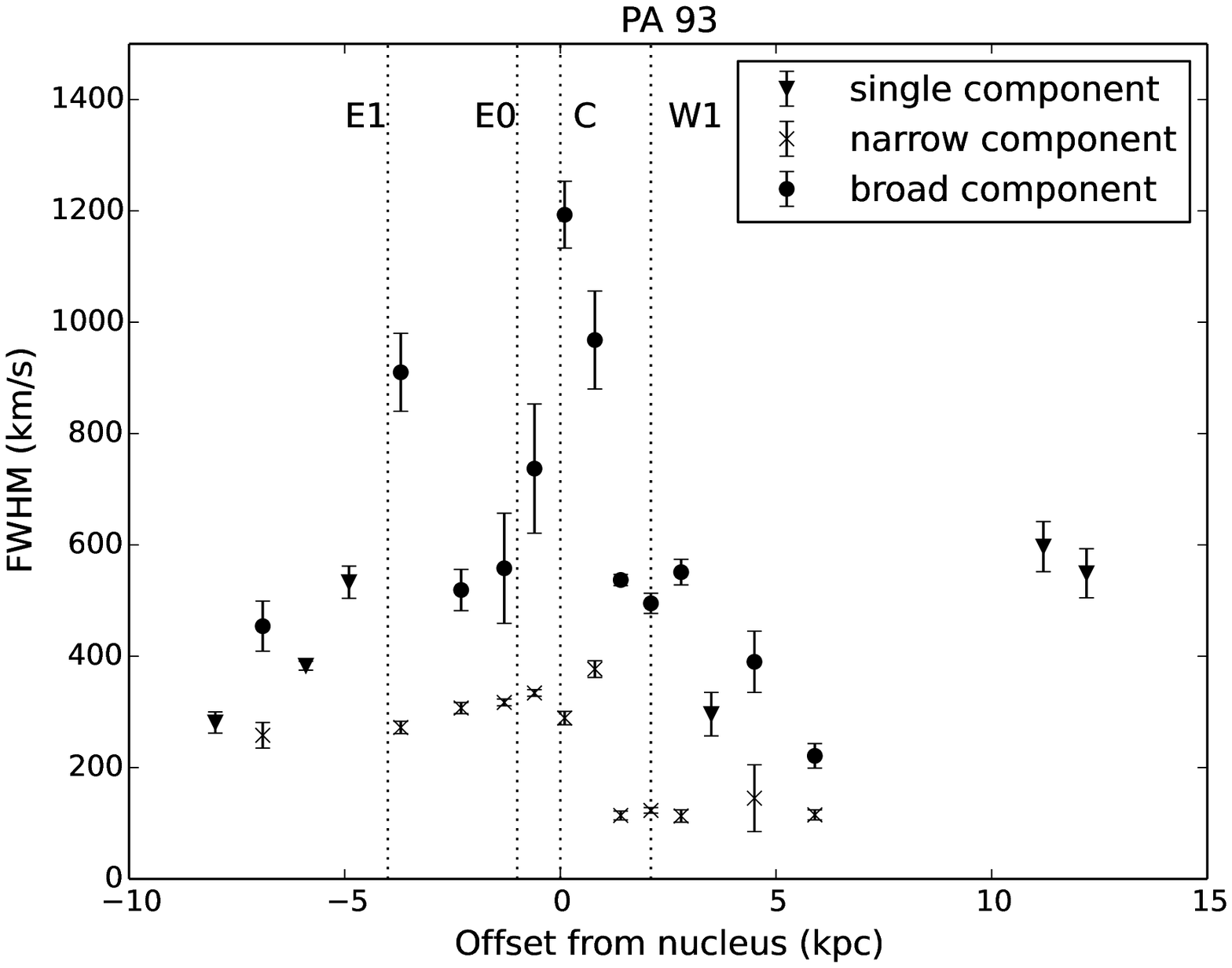, width=\linewidth}}
\end{minipage}
\caption{Line-widths of the gaussian components fitted in the long-slit data as a function of distance from the nucleus of 3C293. The left figure shows the data extracted along the axis of the disk of the galaxy (position angle of 225$^{\circ}$) and the right shows the line-widths measured along the axis of the radio jets (position angle of 93$^{\circ}$). For apertures where a single gaussian component fit was adequate to fit the emission lines the measured line-widths are shown by the triangles. In regions where two gaussian components were fit to the data, these have been separated into narrow components (marked by the crosses) and broad components (circles). For the slit placed along the radio-jet axis we have also marked features of the inner-radio jet structure using the same notation as EM05. The brightest hotspots correspond to E0 and C while E1 and W1 refer to diffuse jet-emission detected beyond the hotspots (see \citealt{Beswick2002}). Positive offsets are west of the nucleus for both slits. \label{disp}}
\end{figure*}

The long-slit data were taken using the Intermediate dispersion Spectrograph and Imaging system (ISIS), mounted on the WHT on the nights of 27 May 2006 and 29 June 2006. These data were taken with the primary aim of measuring the kinematics of the NaI absorption doublet in order to further investigate the neutral gas kinematics (see Bessiere et al., 2016, in preparation), but they are also useful for checking the IFU results on the ionised gas kinematics, since ther spectral resolution is a factor $>$3.5$\times$ higher. ISIS allows for the use of a dichroic filter, thus facilitating simultaneous observing in both the blue and red. The standard 5300 dichroic was used with the R300B grating for the blue arm, the R600R for the red arm and a 1\arcsec slit. We observed 3C293 at two position angles, PA 93 comprises $6 \times 900~\mbox{s}$ exposure whilst PA 225 consists of $6 \times 1200~\mbox{s}$ exposures. The average seeing throughout the the observations made at PA 93 was 0.5\arcsec, whilst those made at PA of 225 had an average seeing of 0.8\arcsec. 

These data was reduced using packages within the {\sc iraf} environment and also the {\sc starlink} package {\sc figaro}. The data were bias subtracted, flat-fielded and wave-length calibrated in the usual manner. The arc exposures were taken immediately before and after the observations at the same positions as the target. To ensure an accurate relative flux calibration between the blue and the red arm, several standard stars were taken on each night and individual flux calibration curves in both the blue and the red were produced for each star. For each night, the flux calibration curves were compared with one another, and any obvious outliers were removed. The individual curves were then combined into a master blue and a master red flux calibration curve for that night. The final reduced data covers the wavelength range 3300 -- 7070\AA\ in the observed frame, with a spectral resolution of 3.7 \AA\ in the blue and 1.8 \AA\ in the red.

Apertures of either 0.2 or 0.4\arcsec were extracted along the slits. The extracted spectrum was then continuum subtracted and Gaussian fits were made using the {\sc starlink dipso} package. The best fit for both the [SII] and the H$\alpha$, [NII] lines were used to determine the velocity shifts and line-widths for each aperture. These are shown in Figure \ref{disp}. In some cases it was possible to fit the profiles with a single Gaussian, but for most of the spectroscopic apertures a double Gaussian was required. Consistent with the results of \citetalias{Emonts2005}, along the inner radio axis (PA93) there is evidence for extreme kinematics in the form of large line widths ($FWHM > 400$ km s$^{-1}$) found out to a radial distance of 7\,kpc to the East and West of the nucleus -- significantly beyond the high-surface-brightness inner radio structures. The results also show evidence that, where double Gaussians were required to fit the lines, the broad components are strongly blue shifted ($\Delta V > 200$~km s$^{-1}$ relative to the narrow components. In addition, these observations, which have a higher sensitivity and spectral resolution than those of EM05, detect a region with disturbed kinematics at a radial distance of 12\,kpc to the west. In this region, the line widths reach $FWHM > 500$~km s$^{-1}$. 

Based on the long-slit data, the evidence for disturbed emission line kinematics is not only confined to the radio axis. The results for PA225 presented in Figure \ref{disp} demonstrate that relatively broad lines ($FWHM > 300$~km s$^{-1}$) are detected out to a radial distance from the nucleus of 3.5\,kpc  to the NE and 5.0 kpc to the SW; at maximum extent, the broad lines are detected out to a distance of 3.5\,kpc in the direction perpendicular to the radio axis\footnote{The line-widths have been corrected for the instrumental width}. Such large line-widths in the off-nuclear regions are inconsistent with the normal gravitation motions of gas in the disk of a galaxy (i.e. \citealt{Labiano2014} show that the kinematics of the inner 7\,kpc CO disk are consistent with rotation around the core). Thus, the long-slit results confirm what was seen in the IFU observations in the sense that they show that the disturbed emission line kinematics have a significant extent perpendicular to the radio axis.

One possible explanation for this is that we are seeing the cocoon structure described in simulations of jet-driven feedback \citep{Wagner2011, Wagner2012}. As the jets propagate through the inhomogenous ISM, they not only impact the ISM along the radio jet axis, but also create a spherical bubble which drives dense clouds outwards in all directions as the bubble expands. Although this mechanism can not impart the gas with the extreme velocities required to escape the galaxy, it does disturb and heat the gas sufficiently to inhibit star formation \citep{Nesvadba2010, Guillard2012}. This turbulent behaviour is similar to what is observed in a growing number of objects exhibiting jet-driven outflows such as IC5063, NGC1266 and 3C326 \citep{Morganti2015, Alatalo2014, Guillard2015}. 

\subsection{Line ratios}

\begin{figure}
\centering{\epsfig{file=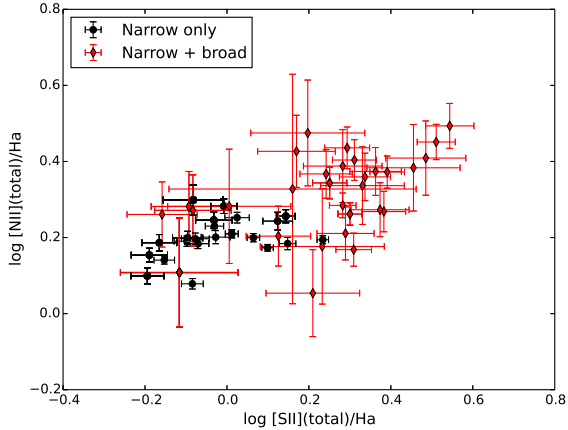, width=\linewidth}}
\caption{Comparison of the [SII]/H$\alpha$ line ratio against the [NII]/H$\alpha$ line ratio.To avoid errors associated with the fitting multiple gaussian components to blended emission lines, we have calculated the ratios using the total [SII]$\lambda$6716+$\lambda$6731 and [NII]$\lambda$6548+$\lambda$6583 fluxes. The line ratios for regions where only narrow components were fitted are shown in the black circles while the red diamonds show the line ratios calculated by adding both the narrow and broad components. \label{lineratio}} 
\end{figure}

Having fit the H$\alpha$, [NII] and [SII] lines simultaneously we can also investigate some line ratio diagnostics. Combining the [NII]/H$\alpha$, [SII]/H$\alpha$ line ratios from the OASIS data with the [OIII]/H$\beta$ ratios from \citetalias{Emonts2005}, we find that 3C293 can be classified as a LINER (see also EM05). The excitation mechanism for LINERs is still not well understood, but in the case of 3C293 our observations can be explained both by dusty AGN and by (pure) shock models \citep{Groves2004,Allen2008}. Given the morphological association of the higher excitation region with the inner radio jets in 3C293 we favour the shock models over the AGN photoionization models. In figure \ref{lineratio} we show the [NII]/H$\alpha$, [SII]/H$\alpha$ line ratios for regions where only a narrow component was fit to the data compared to regions where two gaussian components were needed. There is a clear increase in the line ratios in the regions where we detect outflowing material (i.e. where both a narrow and broad component were fit to the data). In terms of shock models this increase reflects the increase in shock velocity \citep{Dopita1995, Allen2008}. This picture is further supported by the increased line-widths observed in the radio region \citep{Dopita1995}. 

\section{Properties of the outflow} \label{outflow}

\subsection{Geometry of the outflow}

With the addition of the IFU data we can now present a more unified picture of the kinematics of the outflow in 3C293. Previous studies have concluded that, for the inner radio jets, the eastern jet is approaching \citep{Akujor1996, Beswick2004, Floyd2006}. This agrees with the fact that we observe the outflow at the eastern jet to be significantly blueshifted. At the western jet, although the outflow velocities are still blueshifted (as shown in Figure \ref{velmaps}), they are closer to the systemic velocity, similar to the outflow detected in HI \citep{Mahony2013}. Since the western jet is more embedded in the disk of the galaxy it is likely that any redshifted emission associated with this jet is obscured. However, \citetalias{Emonts2005} reported a redshifted outflow component detected in the region beyond the western hotspot consistent with this orientation\footnote{This redshifted emission is also confirmed in the long-slit data presented in Section \ref{longslitsec}}.

\begin{figure}
\includegraphics[width=\linewidth]{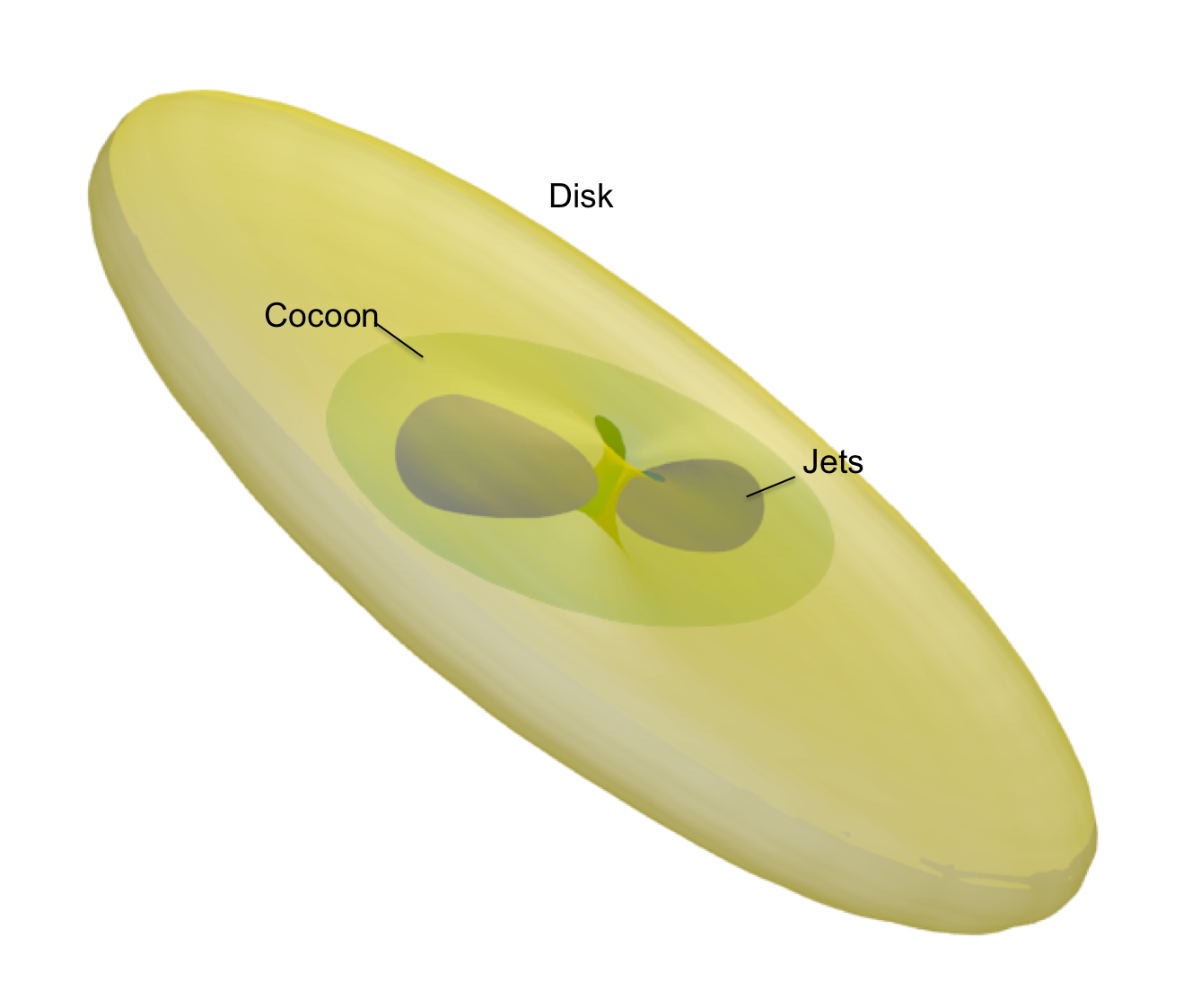}
\caption{Cartoon representation of the orientation and geometry of the nearby radio galaxy 3C293. The disk of the galaxy and the radio jets are roughly orientated the same way (eastern side approaching us) meaning that the jets pass through the galaxy disk. On the eastern side this results is blueshifted outflows as observed in the ionised gas. On the western side we detect broad lines with velocities close to the systemic velocity. We also detect the neutral gas outflow in front of the western radio lobe due to the larger amount of gas along that sightline. As the radio jets propagate through the disk they also create a spherical bubble which disturbs the gas in all directions shown by the cocoon-type structure surrounding the jets. \label{cartoon}}  
\end{figure}

In Figure \ref{cartoon} we show a 3D cartoon representation of what the inner regions of 3C293 may resemble. The disk of the galaxy is orientated such that the eastern side of the disk is towards us \citep{Floyd2006, Labiano2014}. Since the eastern inner radio jet is also approaching this implies that the jet is crossing through the disk. The jets also create a cocoon which disturbs the gas in all directions from the core. 

This orientation can also explain why the neutral gas phase of the outflow is only detected in the western jet. The HI outflows are detected in absorption, meaning that we are only sensitive to the gas that lies along our line of sight in front of the radio continuum. The fact that the outflow is detected all along the radio-jet axis suggests that the outflow is not produced only at the hotspots, but could also indicate that there is fast-moving gas entrained by the radio jets. This, along with the presence of the cocoon, suggests that there is a lot of turbulent gas in the central regions of 3C293. If the western jet is receding, it is likely that there is a much higher column density of gas in front of the jet to detect the broad, shallow component in HI whereas on the eastern side the jet is in front of a lot of this gas. As the western hotspot is close to the core we observe an outflow velocity slightly blueshifted compared to the systemic rather than redshifted as would be the case in more extended radio jets. 

Note that this orientation of the inner radio jets (eastern jet approaching) is the opposite of what is seen in the larger scale radio lobes. There, the north-western lobe is significantly brighter than the south-eastern lobe and therefore assumed to be the approaching jet \citep{Joshi2011}. However, since the orientation of the jet-axis has clearly changed between the outer and inner jets \citep{Beswick2004}, it is possible that the axis has also shifted along the line of sight.

\subsection{Outflow rates}

Having now confirmed that the ionised gas outflow is evident all along the radio jet axis we can calculate more accurately the properties of the outflow. Since the outflow is resolved we calculate the total mass (M$_{gas}$) for the eastern and western jets separately based on the H$\alpha$ luminosities using the following relation (\citetalias{Emonts2005}, \citealt{Osterbrock}):

\begin{equation}
M_{gas} = \frac{m_p L(H\alpha)}{N \alpha^{eff}_{H\alpha} h \nu_{H\alpha}}   .
\end{equation}

Here $m_p$ is the mass of a proton (in kg), $L(H\alpha)$ is the $H\alpha$ luminosity (in ergs\,s$^{-1}$), $N$ the electron density (in cm$^{-3}$), $h$ the planck constant and $\nu_{H\alpha}$ the frequency of $H\alpha$. The effective recombination coefficient, $\alpha^{eff}_{H\alpha}$, is calculated to be $2.87 \cdot 1.24 \times 10^{-25} \slash h \nu_{H\alpha}$ assuming case B recombination theory and a temperature of 10000\,K \citep{Osterbrock}. The density was estimated using the the [SII]$\lambda$6716/$\lambda$6731 ratio, based on single Gaussian fits to each of the two lines in the [SII] blend in each of the spectroscopic apertures of the long-slit data\footnote{Note that there was not sufficient S/N to allow the [SII] ratio to be determined for broad and narrow components separately. Therefore the estimated density is an average for both components.}. Significant density measurements (i.e. the [SII] ratio measurement was $> 3\sigma$ from the low density limit) were obtained in two apertures: one 1.4 arcseconds NE of the nucleus along PA225 ($N_e = 368 \pm 107$ cm$^{-3}$), and the other 1.4 arcseconds to the E of the nucleus along PA93 ($N_e = 110 \pm 31$ cm$^{-3}$). However, all the [SII] ratio estimates are consistent within the errors for all the spectroscopic apertures. Therefore we adopted the density derived\footnote{assuming a temperature of 10,000K} from the error weighted mean of all the [SII] estimates ($N_e = 204\substack{+49 \\ -44}$\,cm$^{-3}$) as a representative density for all the apertures.

Using H$\alpha$ luminosities of $L(H\alpha)=3.7 \times 10^{40}$ for the eastern hotspot and $L(H\alpha)=2.5 \times 10^{40}$\,erg\,s$^{-1}$ for the western hotspot, we calculate the total mass of ionised gas to be $2.9 \times 10^{5}$ and $4.4 \times 10^{5}$ \mdot  for the eastern and western jets respectively.

To calculate the mass outflow rate associated with each hotspot we used the measured outflow velocites of $v_{out}=304$\,km\,s$^{-1}$ and $v_{out}=90$\,km\,s$^{-1}$ and divided this by the width of each bin in the direction along the jet (0.8 and 0.5\,kpc respectively) to derive the time taken for gas to travel that distance. This gives mass outflow rates of 0.17 \mdot\,yr$^{-1}$ for the eastern jet and 0.05 \mdot\,yr$^{-1}$ for the western jet. From these mass outflow rates we can then calculate the kinetic power, including both the radial and turbulent components of the gas, from the following relation \citep{Holt2006}: 

\begin{equation}
\dot{E} = 6.34 \times 10^{35} \frac{\dot{M}}{2} \left(v^{2}_{out} +  3\sigma^{2}\right) {\rm erg}\,{\rm s}^{-1} 
\end{equation}

where $\sigma=$FWHM$/2.35$. Using FWHM values of 1750 and 1250 km\,s$^{-1}$ for the eastern and western outflows respectively, this gives us kinetic energy rates ranging from $4.6\times 10^{39}$\,erg\,s$^{-1} - 3.5\times 10^{40}$\,erg\,s$^{-1}$. This represents only a small fraction of the Eddington luminosity ($\dot{E}\slash L_{edd} \sim 0.4-3\times 10^{-6}$) and is significantly less than the associated outflow of neutral gas (8--50 \mdot\,yr$^{-1}$), even though this was only detected in front of one of the radio jets \citep{Mahony2013}. This is similar to what is found in other sources where the majority of the mass is in the neutral and molecular gas phases and only a small fraction of the total mass outflow rate is detected in the warm, ionised gas (i.e. \citealt{Feruglio2010, Dasyra2012, Morganti2013a, Garcia-Burillo2014}). Combining the total energy rate for both the ionised and neutral gas we calculate the kinetic power injected into the ISM to be of order 0.01--0.08 per cent of the Eddington luminosity \citep{Mahony2013}.

\section{Conclusions} \label{concl}

In this paper we presented IFU observations of the nearby radio galaxy 3C293 observed with OASIS on the WHT. Previous long-slit observations of this galaxy had detected outflows of ionised gas associated with the eastern inner radio lobe \citepalias{Emonts2005}, but HI absorption studies had located an outflow of neutral gas in front of the western radio lobe \citep{Mahony2013}. The IFU observations allowed us to spatially map out the outflow and resolve these seemingly contradictory results. 

To measure the kinematics of the gas, either 1 or 2 gaussian components were fit to each spectrum. Where 2 gaussian components were needed to fit the data, the lines were separated into a `narrow' and `broad' component representing the gas associated with the disk of the galaxy and the outflow respectively. Using this method we detected an outflow of ionised gas all along the radio jet axis. In addition to detecting fast outflows associated with the radio jet, we also detected broad line-widths (up to 500\,km\,s$^{-1}$) in regions up to 12\,kpc from the nucleus. We suggest that this could be evidence of the cocoon structure seen in simulations of \citet{Wagner2011, Wagner2012}, which show that as the jet propogates through a clumpy ISM it also inflates a spherical-like bubble capable of disturbing the gas in all directions. In the case of 3C293, this region of disturbed gas extends up to 12\,kpc from the nucleus along the radio-jet axis and up to 3.5\,kpc in the direction perpendicular to the jets.  

We calculate mass outflow rates of $0.05-0.17$ \mdot\,yr$^{-1}$ and corresponding kinetic powers of $\sim 0.5-3.5\times 10^{40}$\,erg\,s$^{-1}$. This represents a tiny fraction of the Eddington luminosity ($\dot{E}\slash L_{edd} \sim 1\times 10^{-6}$), two orders of magnitude less than the kinetic power of the corresponding neutral gas outflow. 

To further investigate the detailed kinematics of the central regions of 3C293, higher spatial resolution data are required to further map the cocoon structure. Higher S/N across a larger area would also allow us to more readily detect the faint, broad wings indicative of outflows. In addition, IFU observations covering the H$\beta$ and [OIII] lines would enable us to study the emission line diagnostics to better understand the properties of both the regularly rotating gas and the outflow. Finally, studying the molecular gas properties of the outflow with deeper CO observations at higher spatial resolution would provide a comprehensive view of the jet-ISM interaction in 3C293.

\section*{Acknowledgements}

The William Herschel Telescope is operated on the island of La Palma by the Isaac Newton Group in the Spanish Observatorio del Roque de los Muchachos of the Instituto de Astrofísica de Canarias. We thank Lilian Dominguez for her support when carrying out the observations and data reduction, Tom McCavana for his help in creating Figure \ref{cartoon} and the anonymous referee for their useful comments. The research leading to these results has received funding from the European Research Council under the European Union's Seventh Framework Programme (FP/2007-2013) / ERC Advanced Grant RADIOLIFE-320745. BE acknowledges funding through the European Union FP7-PEOPLE-2013-IEF grant nr. 624351.

\bibliographystyle{mn2e}
\setlength{\bibhang}{2.0em}
\setlength\labelwidth{0.0em}

\bibliography{3c293}

\bsp

\label{lastpage}

\end{document}